\documentclass[aps,prb,preprint,groupedaddress,showpacs]{revtex4}
\usepackage{graphicx}
\usepackage{amssymb}
\usepackage{amsmath}
\usepackage{latexsym}
\usepackage{ulem}
\usepackage{color}
\usepackage{dcolumn}
\usepackage{subfigure}

\usepackage[T1]{fontenc}
\usepackage[utf8]{inputenc}

\usepackage[english,frenchb]{babel}

\begin{document}

\title{Unraveling the nature of carrier mediated ferromagnetism in diluted magnetic semiconductors}
\author{Georges Bouzerar and Richard Bouzerar}
\affiliation{
Institut Lumi\`ere Mati\`ere, CNRS  et Universit\'e  Lyon 1 \\
6, rue Ada Byron \\
69622 Villeurbanne Cedex, France
}

\date{\today}

\clearpage
\selectlanguage{english}
\begin{abstract}
After more than a decade of intensive research in the field of diluted magnetic semiconductors (DMS),  the nature and origin of ferromagnetism, especially in III-V compounds is still controversial.
Many questions and open issues are under intensive debates. 
Why after so many years of investigations Mn doped GaAs remains the candidate with the highest Curie temperature among the broad family of III-V materials doped with transition metal (TM) impurities ?
How can one understand that these temperatures are almost two orders of magnitude larger than that of hole doped (Zn,Mn)Te or (Cd,Mn)Se? Is there any intrinsic limitation or is there any hope to reach in the dilute regime room temperature ferromagnetism? How can one explain the proximity of (Ga,Mn)As to the metal-insulator transition and the change from Ruderman-Kittel-Kasuya-Yosida (RKKY) couplings in II-VI compounds to double exchange type in (Ga,Mn)N?
In spite of the great success of density functional theory based studies to provide accurately the critical temperatures in various compounds, till very lately a theory that provides a coherent picture and understanding of the underlying physics was still missing. Recently, within a minimal model it has been possible to show that among the physical parameters, the key one is the position of the TM acceptor level. By tuning the value of that parameter, one is able to explain quantitatively both magnetic and transport  properties in a broad family of DMS. We will see that this minimal model explains in particular the RKKY nature of the exchange in (Zn,Mn)Te/(Cd,Mn)Te and the double exchange type in (Ga,Mn)N and simultaneously the reason why (Ga,Mn)As exhibits the highest critical temperature among both II-VI and III-V DMS's.
\end{abstract}

\pacs{}
\maketitle

\selectlanguage{francais}
{\bf \large Résumé}

Après plus d’une décennie de recherches intensives dans le domaine des semi-conducteurs magnétiques dilués (DMS), la nature et l'origine du ferromagnétisme, en particulier dans les composés III-V, restent  controversées. De nombreuses questions  et problèmes ouverts sont toujours sujets à d’intenses débats. Pourquoi parmi la grande famille des matériaux III-V, et pour une concentration donnée en métal de transition, le composé (Ga,Mn)As reste-t-il le candidat présentant encore la température critique la plus élevée? Comment peut-on comprendre que ces températures soient presque deux ordres de grandeur supérieures à celles observées dans (Zn,Mn)Te dopé en trous ou (Cd,Mn)Se?
Subsiste-t-il pour ces matériaux dilués un espoir d’observer un ordre ferromagnétique au delà de la température ambiante ou est-il fatalement anéanti par des limitations physiques intrinsèques ? 
Comment expliquer que (Ga,Mn)As soit si proche de la transition métal-isolant ? Comment comprendre la nature des couplages magnétiques passant typiquement de RKKY dans les composés II-VI à double échange dans (Ga,Mn)N?
Des études, basées sur la théorie de la fonctionnelle de la densité, ont pu fournir avec précision les températures critiques dans divers composés.  Cependant un modèle théorique en mesure de fournir une vision unifiée et une compréhension de la physique sous-jacente manquait toujours. Très récemment, dans le cadre d’un modèle minimal, il a été possible de montrer que, parmi les paramètres physiques, la clé réside dans la position du niveau accepteur de l’impureté magnétique. En adaptant ce dernier, il devient en effet possible d'appréhender la diversité des propriétés magnétiques et aussi de transport dans une large famille de DMS.
Nous verrons alors que le modèle minimal explique non seulement la nature RKKY des couplages magnétiques dans (Zn,Mn)Te/(Cd,Mn)Te ou leur caractère double échange dans (Ga,Mn)N, mais aussi la raison pour laquelle (Ga,Mn)As présente les températures de Curie les plus élevées parmi les DMS II-VI et III-V. 

\selectlanguage{english}

\section{Introduction}
\begin{figure}[t]\centerline
{\includegraphics[width=5.0in,angle=0]{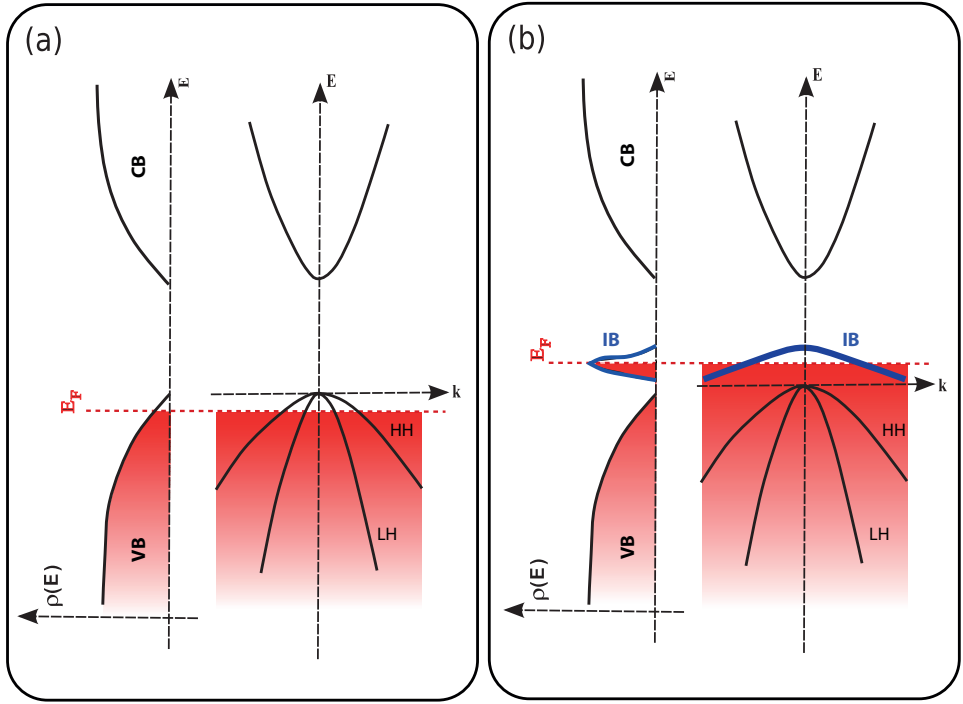}}
\caption{(Color online) Schematic view of the two opposite scenari for the physics in the III-V Mn doped semiconductor GaAs: (a) Valence band picture (VB) and (b) Impurity Band scenario (IB). LH and HH are respectively the light and heavy hole band near the center of the Brillouin zone.
}
\label{fig1}
\end{figure}

\ In recent  years, the rapidly growing field of diluted magnetic semiconductors (DMS) \cite{junqwirth,satormp,timm} has attracted a considerable interest owing to their potential for spintronic devices. One of the main goals is to combine,  the traditional electronic functionality (charge) and the spin degree of freedom of  the electrons/holes. This requires optimal candidates that exhibit room temperature ferromagnetism. Recent progress in growth processes of TM-doped III-V semiconductors  has boosted the interest for such novel materials. Among III-V DMS, Mn doped GaAs that could be considered as the prototype is certainly the most widely studied (both transport and magnetic properties). However, the understanding of the fundamental physical properties in these doped compounds involves  theoretical speculations which are subject to controversy. The quest for a model able to capture quantitatively the physics and identify the key physical parameters that control both magnetic and transport properties  was a clear open issue over the last decade.
Till recently DMS based theoretical studies could be split into two main distinct types: (i) First principle based approaches \cite{satormp} and (ii) Zener Mean Field type theories (ZMF) \cite{dietl,junqwirth}. The first kind is based on density functional theory (DFT) such as local spin density approximation (LSDA) or generalized gradient approximation (GGA) for instance. They require no adjustable parameters and are essentially material specific. The second type is a model approach that includes a realistic description of the host band structure within a 6 bands or 8 bands Kohn-Luttinger Hamiltonian \cite{kl,kane} and a local p-d exchange between itinerant holes and localized impurity spins. In Zener Mean Field theory the p-d coupling is treated pertubatively and the dilution effects at the lowest order, also known as Virtual Crystal Approximation (VCA). As a consequence, the Fermi level lies inside the unperturbed valence band (VB) leading to the so called valence band scenario (see Fig.~\ref{fig1}(a)).
Regarding the specific case of (Ga,Mn)As, the perturbative VB picture is inconsistent with first principle based studies. Indeed, density functional calculations clearly predict the existence of a well defined preformed impurity band (see  Fig.~\ref{fig1}(b)) that for a sufficiently large concentration of Mn (beyond 1\%) overlaps with the Valence band \cite{satormp,wierzbo}. It is found that the Fermi level lies in the resonant impurity band. Thus, ab initio studies clearly support the so called "impurity band picture" (IB). It is worth noticing that both  optical conductivity measurements \cite{burch,singley1,singley2} and proximity of Mn doped GaAs to the Metal-Insulator transition \cite{hayashi,matsukura,yazdani} fully support the IB-picture. In spite of all that the issue of VB versus IB scenario is still controversial. On the other hand, the VB scenario remains suitable to describe the physics in II-VI materials such as Mn doped ZnTe, CdTe, ZnSe for instance. 
The reason for this is the absence of hybridized p-d states in the vicinity of the top of the valence band in these alloys. In other words, treating the substitution by Mn as a perturbation remains a good approximation in II-VI materials.
In the following, we present a two step approach that allows to describe both magnetic and/or transport properties of a wide range of diluted magnetic semiconductors. Concerning the magnetic properties that are our main concern in this paper the two steps are described as follows. The first one consists in calculating the magnetic couplings between localized spins randomly distributed in the semiconductor host. To this end one can use first principle calculations or suitable model approaches. The purpose is to build  the effective Heisenberg Hamiltonian of the problem. This spin Hamiltonian is diagonalized during the second step within the self consistent local random phase approximation \cite{georges2005epl} (SC-LRPA) procedure that is described in the next section.

\section{Self-consistent Local RPA diagonalization of the dilute Heiseinberg Hamiltonian}

The Hamiltonian that describes N$_{\rm imp}$ interacting spins ${\bf S}_{i}$  (classical or quantum) randomly distributed in the host lattice is the dilute/disordered Heisenberg model,
\begin{eqnarray}
H_{\rm Heis}=-\sum_{i,j} J_{ij}p_{i}p_{j} {\bf S}_{i}\cdot{\bf S}_{j}
\label{Hamiltonian-Heisenberg}
\end{eqnarray}
$J_{ij}$ are the spin-spin couplings, the sum runs over all sites and the random variable p$_i$ is 1 if the site is occupied by an impurity, otherwise 0. In the case of Mn doped III-V DMS, the localized spin S=5/2, thus it can be treated classically.

In what follows the dilute Heisenberg Hamiltonian is diagonalized using the Self-Consistent Local Random Phase Approximation (SC-LRPA).  It is a semi-analytical method based on finite temperature Green's functions that describe the spin fluctuations. This powerful approach offers several advantages. Compared to standard classical Monte Carlo simulations (MC), (i) it allows calculations on large system sizes, (ii) the CPU and memory cost are relatively low, (iii) the critical temperature is given by a semi-analytical expression (no need to calculate Binder cumulants as in MC), (iv) the T-dependent local magnetizations, susceptibility and magnetic excitation spectrum can be calculated directly as well. 

In order to calculate the magnetic properties we define the following retarded Green's function $G^{S}_{ij}(\omega)=\int_{-\infty}^{+\infty}G^S_{ij}(t)e^{i\omega t}dt$, where 
$G^S_{ij}(t)=-i\theta(t)\langle[S_i^+(t),S_j^-(0)]\rangle$, ($\langle ...\rangle$ denotes the thermal average). G$^{S}_{ij}$ describes the transverse and thermal spin fluctuations. After Tyablicov (or RPA) decoupling of the higher-order Green's functions that appear in the equation of motion of  $G^S_{ij}(\omega)$ one finds,
\begin{eqnarray}
G_{ij}^S(E)=2 \lambda_{j} \langle i| \frac{1} {E-H_{\rm eff}+i\epsilon} |j\rangle
\label{GF}
\end{eqnarray}
For convenience we have introduced the reduced variable $E=\omega/m$, $m$ being the averaged magnetization in the sample ($m=\frac{1} {\rm N_{imp}} \sum_{j} \langle{S_j^z}\rangle$), and 
$\lambda_{j} =\frac{\langle{S_j^z}\rangle} {m}$. For each configuration of the disorder (random positions of the impurities), the local T-dependent magnetizations $\{\langle{S_{j}^z}\rangle_{j=1,2,...,N_{imp}}\}$ are calculated self-consistently.

 ${\bf H}_{\rm eff}$ is an effective  N$_{\rm imp}\times$N$_{\rm imp}$  non Hermitian, bi-orthogonal matrix \cite{dieudonne1953} whose matrix elements are,
\begin{eqnarray}
({{\bf H}_{ \rm eff})}_{ij}=- \lambda_{i}J_{ij}+ \delta_{ij}\sum_{l}  \lambda_{l}  J_{lj}
\label{heff}
\end{eqnarray}
The property of bi-orthogonality implies a set of right and left eigenvectors $|\Psi_\alpha^{R}\rangle$ and $|\Psi_\alpha^{L}\rangle$, each pair associated with the same eigenvalue $E_{\alpha}$ ($\alpha$=1,2,..N$_{\rm imp}$).

The retarded Green's function can be re-written,
\begin{eqnarray}
G_{ij}^S(E)=2 \lambda_{j}  \sum_\alpha \frac{\langle i|\Psi_\alpha^{R}\rangle \langle\Psi_\alpha^{L}|j\rangle}{E-E_\alpha+i\epsilon}
\end{eqnarray}

\begin{figure}[t]\centerline
{\includegraphics[width=5.0in,angle=0]{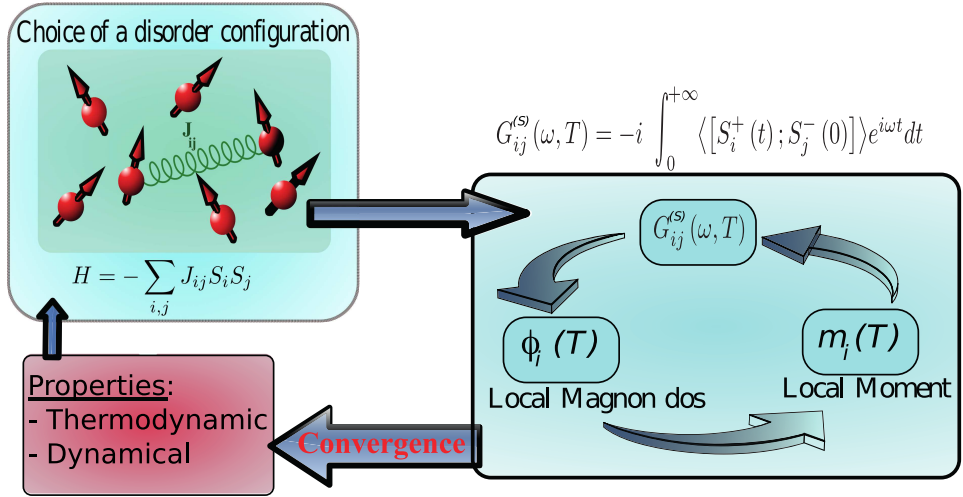}}
\caption{(Color online) Illustration of the Self Consistent-Local RPA loop used to diagonalize the dilute/disordered Heisenberg Hamiltonian (see ref. \onlinecite{georges2005epl} for details).
}
\label{fig2}
\end{figure}

The SC-LRPA has been used in several studies and was proven accurate and reliable. It properly treats thermal fluctuations and disorder effects such as localization and percolation physics.
One can for instance find detailed discussions in Refs \onlinecite{georges2005epl,satormp,georges2007,akash2010}. A summary of the SC-LRPA procedure is illustrated in Fig.~\ref{fig2}. As mentioned previously, one of the great advantages of the SC-LRPA is to allow a direct calculation of the critical temperature. Indeed, one can derive a semi-analytical expression for T$_C$ that reads \cite{georges2005epl},
\begin{eqnarray}
k_{B} T_{C} =\frac{2}{3} S(S+1)  \langle \frac{1}{F_{i}}\rangle
\label{tc1}
\end{eqnarray}
where $ \langle F_{i}^{-1}\rangle = \frac{1}{N_{\rm imp}} \sum_{i} F_{i}^{-1} $ and 
\begin{eqnarray}
F_{i} =-\frac{1}{2\pi \lambda_{i}}\int_{-\infty}^{+\infty} \frac{\rm Im G^{S}_{ii}(E)}{E}dE=\sum_\alpha \frac{\langle i|\Psi_\alpha^{R}\rangle \langle\Psi_\alpha^{L}|i\rangle}{E_\alpha}
\label{tc2}
\end{eqnarray}
 F$_{i}$'s depend on the reduced local magnetizations $\{ \lambda_{j} \}_{j=1,2,...,N_{imp}}$ that are calculated self-consistently. In the limit of vanishing magnetization one finds the following set of equations,
\begin{eqnarray}
\lambda_{j} (T \rightarrow T_{C}) =\frac{F_{j}^{-1}}{\sum_{i} F_{i}^{-1}} 
\end{eqnarray}
The nature of the magnon modes (localized/extended) is explicitly taken into account in the expression of T$_{C}$ (eq.(\ref {tc1}) and (\ref {tc2})). The accuracy and reliability of the SC-LRPA to handle both thermal fluctuations and disorder (localization, percolation) has been often addressed by direct comparison with Monte Carlo calculations \cite{georges2005epl,gb-prb-compens,rb-prb-compens, gb-prb-canting,gb-de-2007}. The agreement was systematically very good, the critical temperatures calculated by both methods usually differ by less than 10$\%$. For instance, in the case of 5$\%$ doped GaAs, using the same couplings, Monte Carlo calculations have lead respectively  to 137 K \cite{MC1} and 110 K \cite{MC2} (error bars were not given but should be at least of the order of 10$\%$), whilst the SC-LRPA value is 132 $\pm 5$  K. 
The use of the two step approach combining first principle calculations of the magnetic couplings and SC-LRPA treatment of the effective diluted Heisenberg Hamiltonian is a tool of choice. However, due to the complexity and material specific nature of first principle calculations, it remains difficult to discern the relevant parameters that govern the physical features in various diluted magnetic semiconductors. A suitable minimal model allowing a coherent and consistent picture of the physics in these materials is needed. In the next section we introduce such a model, the V-J Hamiltonian. It
captures most of the relevant features in dilute magnets. It continuously shows how the couplings change from RKKY nature in II-VI Mn doped compounds such as (Zn,Mn)Te or (Cd,Mn)Te to double exchange type in (Ga,Mn)N. We will see that (Ga,Mn)As is located near the metal to insulator transition and the resonant effects due to the position of the Mn acceptor level is responsible for the highest Curie temperature observed in this family of materials.

 \section{Ab initio vs V-J Model studies of the magnetic properties of $\rm \bf Mn$ doped III-V compounds}

The aim of this section is to compare (i) ab initio, (ii) model based calculations and (iii) experimental data. It will be seen that the non perturbative V-J model \cite{richard-epl2007,richard1} provides naturally a coherent and unified picture of the physics (magnetism and transport) in DMS. As mentioned in the introduction part, in Zener Mean Field theory the focus is put on the realistic description of the host band structure (Kohn-Luttinger Hamiltonian) whilst the coupling between holes and localized spins (S=5/2) of Mn$^{2+}$ is treated perturbatively. 
In contrast, the key point of the V-J model (defined below) lies in the non-perturbative treatment of the impurity-hole coupling, while the details in the host band structure are ignored. As will be shown, it appears that the details of the band structure indeed play a secondary role. Note, that the V-J model has been recently extended by including a more realistic description of the host band structure. This study has given further support to the validity of the model\cite{barthel}. 

\begin{figure}[t]\centerline
{\includegraphics[width=5.00in,angle=0]{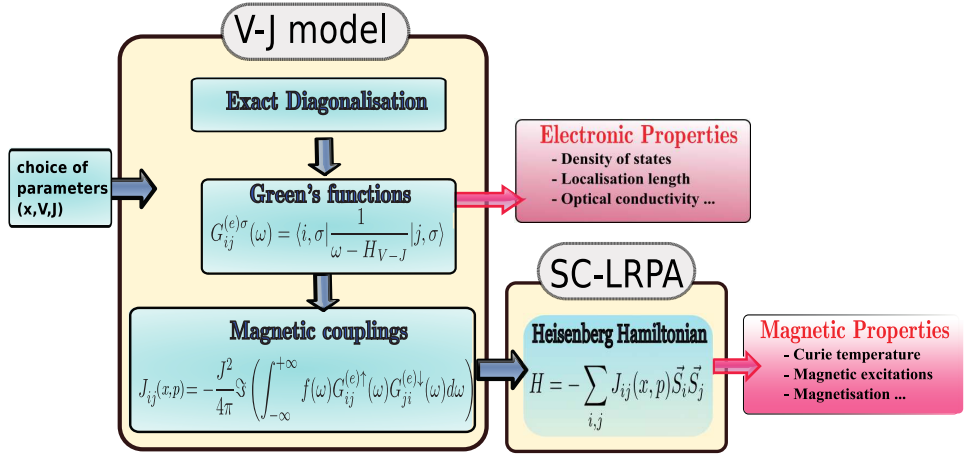}}
\caption{(Color online) Two steps approach used to calculate  both transport and magnetic properties within the V-J model. For a given set of parameters, this procedure is repeated for a sufficiently large number of disorder configurations. $x$ denotes the impurity density and $p$ is the carrier concentration.
}
\label{fig3}
\end{figure} 
The  V-J model is defined as follows\cite{richard-epl2007,richard1},
\begin{eqnarray}
H_{\rm V-J}=-\sum_{i,j,\sigma} t_{ij} c^{\dagger}_{i\sigma}  c_{j\sigma} + \sum_{i}  J_{i} {\bf S}_{i}\cdot{\bf s}_{i} + \sum_{i\sigma}  V_{i}c^{\dagger}_{i\sigma}  c_{i\sigma}
\label{Hamiltonian-VJ}
\end{eqnarray}
where the hopping term t$_{ij}=t$ for i and j nearest neighbours, otherwise zero. $c^{\dagger}_{i\sigma}$  ($c_{i\sigma}$) is the creation (annihilation) operator of a hole of spin $\sigma$ at site i.
 J$_{i}$=J if the site is occupied by Mn otherwise it is zero.  J is the p-d coupling between the localized Mn spin ${\bf S}_{i}$ (S=5/2) and the itinerant hole quantum spin ${\bf s}_{i}$. The on-site potential V$_{i}$ results from the substitution of the host cation by the magnetic impurity. Thus J$_{i}$=p$_{i}$J and V$_{i}$=p$_{i}$V where p$_{i}$=1 if the site is occupied by an impurity, otherwise zero. The one band model contains 3 parameters only (t,J,V). The hopping term has been set to t=0.7 eV (to reproduce the typical VB bandwidth in III-V/II-VI semiconductors).  J is of the order of 1 eV in both Mn doped II-VI and III-V DMS, thus, J has been set to 1.2 eV (widely accepted value for Mn doped GaAs). In this way, the remaining last parameter V fully characterizes a given Mn doped compound. It is chosen in order to reproduce the specific position of the acceptor hybridized p-d state\cite{richard1}. We will see that the on-site scattering potential V, missing in previous theories  \cite{dietl,junqwirth}, is the key parameter. This is discussed in what follows. Fig.~\ref{fig3} shows an illustration of the two steps procedure used to calculate  both transport and magnetic properties within the V-J model. 
 During the first step, the Hamiltonian is diagonalized exactly for each disorder configuration (random positions of the magnetic impurities) in both spin sectors . Then, from the itinerant carrier Green's functions $G_{ij}^{\sigma}$, the magnetic couplings between localized impurity spins are calculated for all distances (see ref. \onlinecite{richard-epl2007} for further details). Note that transport properties can be computed at this stage.  Next, the calculated magnetic couplings enter the Heisenberg Hamiltonian which is treated in the second step in the framework of the SC-LRPA.
In this paper, we have chosen to restrict ourselves to magnetic properties only. Moreover, we will not mention here the effects of compensation defects. This issue has been addressed in several papers\cite{gb-prb-compens,rb-prb-compens,gb-prb-canting}. Transport properties in the framework of the V-J model (Metal-Insulator phase transition and optical conductivity) have been discussed in Ref. \onlinecite{transport}. We would like to stress that a good quantitative agreement between this theory and experiments has been found for transport properties as well. 

\begin{figure}[t]\centerline
{\includegraphics[width=5.0in,angle=0]{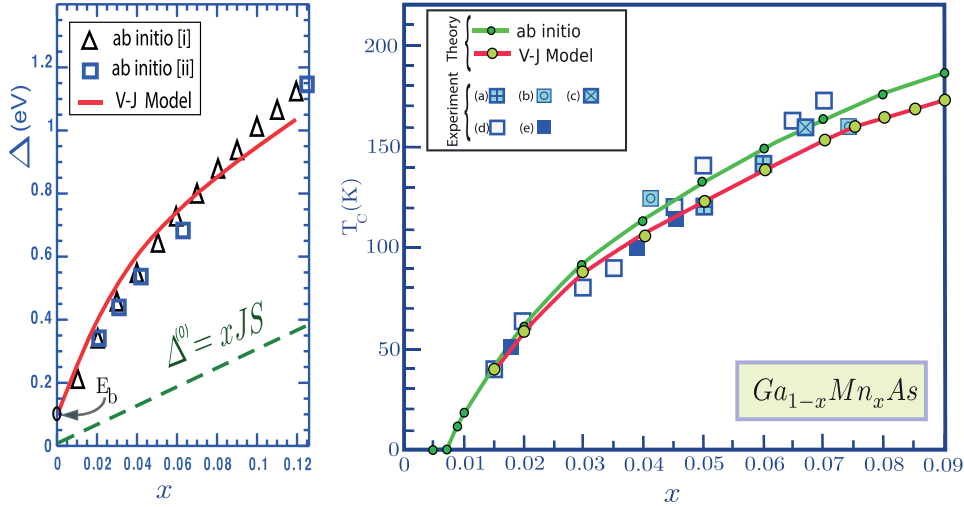}}
\caption{
(Color online) 
(Left) Spin splitting $\Delta$ (eV) as a function of $x$ in Mn doped GaAs (Ab initio and V-J model).  $\Delta^{(0)}$ denotes the pertubative value. The calculations are extracted from Ref.\onlinecite{richard1}. (Right) Measured and calculated (Ab initio and V-J model) critical temperatures in Ga$_{1-x}$Mn$_{x}$As as a function of $x$. The calculations are performed for well annealed compounds (no compensation defects, e.g. 1 hole/Mn).  The experimental data are from (a) Edmonds et al.  \cite{edmonds1}, (b) Chiba et al. \cite{chiba}, (c) Edmonds et al. \cite{edmonds2}, (d) Jungwirth et al. \cite{jungwirth}, (e) Stone et al.  \cite{stone}.
}
\label{fig4}
\end{figure} 

In Fig.~\ref{fig4} we focus first on the case of Mn doped GaAs. The parameter V has been set to 1.8 t=1.26 eV in order to reproduce the position of the measured p-d acceptor level (110 meV). We clearly observe in Fig.~\ref{fig4}-left a very good agreement between V-J model and ab initio calculated spin splitting of the valence band up to 12$\%$ doped samples. On the other hand, the perturbative estimate (Zener Mean Field value) of the spin splitting ($\Delta^{(0)} (x)=x$JS) is found much smaller than that obtained from ab initio studies (about 4 times smaller for the 5$\%$ doped sample). It is also worth mentioning that in the absence of V (V=0), $\Delta$ is well approximated by $\Delta^{(0)}$. This regime corresponds to that of Mn doped II-VI compounds such as Zn$_{1-x}$Mn$_{x}$Te, Cd$_{1-x}$Mn$_{x}$Se for instance. In Fig.~\ref{fig4}-right, we compare the theoretical values of the Curie temperature (ab initio and V-J model) to available experimental measurements. First, it is seen that ab initio and V-J model calculations agree surprisingly well with each other. In addition, we clearly find an overall very good agreement between theory and experiments in the whole concentration range. Note that much more experimental data can be found in the literature for T$_C$. We have only chosen some values corresponding to annealed samples. Because of the presence of compensating defects, as grown samples usually exhibit much smaller critical temperatures. This issue has been theoretically addressed in other papers \cite{gb-prb-compens,rb-prb-compens,gb-prb-canting}. Thus, an excellent quantitative agreement is found between V-J model and ab initio on one side and with the experimental data on the other side. This clearly supports the fact (i) that the V-J model captures the relevant and essential physics and (ii) that the key physical parameter is indeed the acceptor level position.

\begin{figure}[t]\centerline
{\includegraphics[width=5.0in,angle=0]{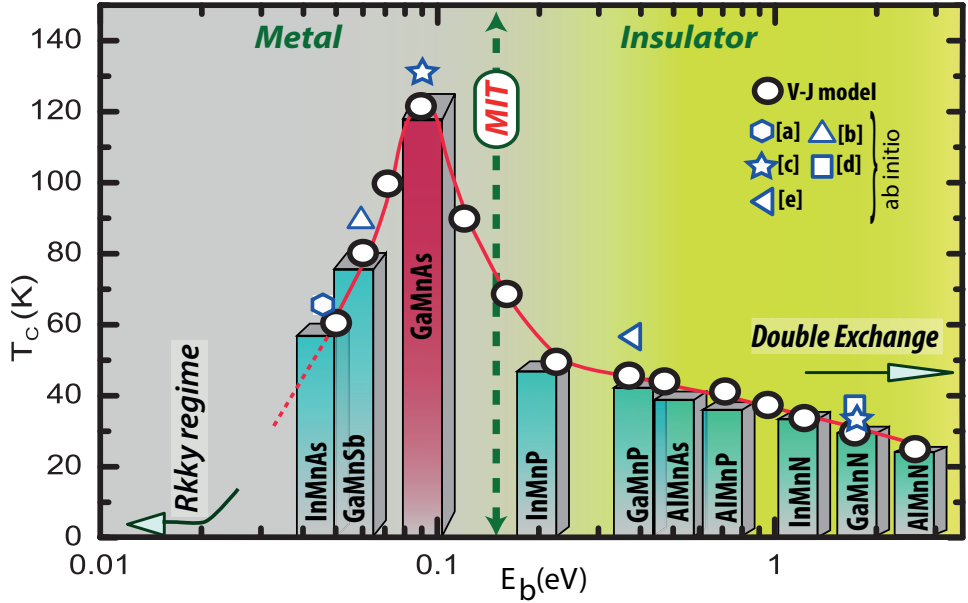}}
\caption{(Color online) Curie temperature (in K) as a function of the acceptor level position E$_b$ (eV). The Curie temperatures result from a systematic average over a large number of disorder configurations. The calculations are systematically performed on large systems (negligible finite size effects).
The concentration of Mn is fixed $x=0.05$ and the hole concentration is $p=x$ (well annealed systems). The open circles are obtained with the V-J model and the other symbols ((a)-(e)) to T$_C$ computed with ab initio exchange integrals  (see Ref. \onlinecite{richard1}). The dashed vertical line correspond to the calculated metal-insulator transition.
}
\label{fig5}
\end{figure} 
We now proceed further and show that the agreement between V-J model and ab initio is not limited to the case of Mn doped GaAs only. For that purpose we have calculated T$_C$ as a function of the acceptor level E$_b$ (by tuning V) and compared it with existing ab initio calculations. The concentration of Mn is set to $x$=0.05 and we have considered the case of optimally annealed systems, thus the hole density is set to $p=x$. The results are depicted in Fig.~\ref{fig5}.  Regarding the ab initio results, the  $x$-coordinate is the measured or calculated Mn acceptor level in the compound.
We observe that the V-J calculated T$_C$ increases rapidly with increasing E$_b$ till it reaches a maximum and then decreases. Note that the rapid increase of T$_C$ occurs on a very short energy scale.  After the maximum, we observe two regimes: first T$_C$ decreases rapidly till E$_b$ $\approx 0.3 ~eV$  and then the decay slope becomes significantly reduced. The maximum of the Curie temperature which is about 125$~K$ is reached for E$_b$ $\approx 0.1~eV$. Remarkably, this acceptor level energy coincides almost exactly with that of Mn doped GaAs (measured value is 110 ~meV). Thus the V-J model explains for the first time why among II-VI and III-V magnetic impurity doped semiconductors the highest measured T$_C$ is that of Mn doped GaAs. 
In this figure, we also see the very good quantitative agreement between the V-J calculated critical temperatures and that obtained from ab initio based calculations.
The reason of this maximum is the fact that the couplings are optimal (resonant effects) when the acceptor level is not too far and not too close to the top of the valence band. When the acceptor level is too small or vanishes (E$_b$ $\rightarrow$ 0) the couplings are RKKY like (case of Mn doped II-VI ) this leads to small T$_C$ (1-2 K) or eventually spin glass phase because of the frustration effects. As we increase  E$_b$ the couplings rapidly loose their oscillating character and become more and more ferromagnetic, thus T$_C$ increases until it reaches its maximum. Afterwards, when the acceptor level becomes larger and larger the ferromagnetic couplings becomes shorter range (double exchange regime), thus leading to a suppression of ferromagnetism.

\begin{figure}[t]\centerline
{\includegraphics[width=6.50in,angle=0]{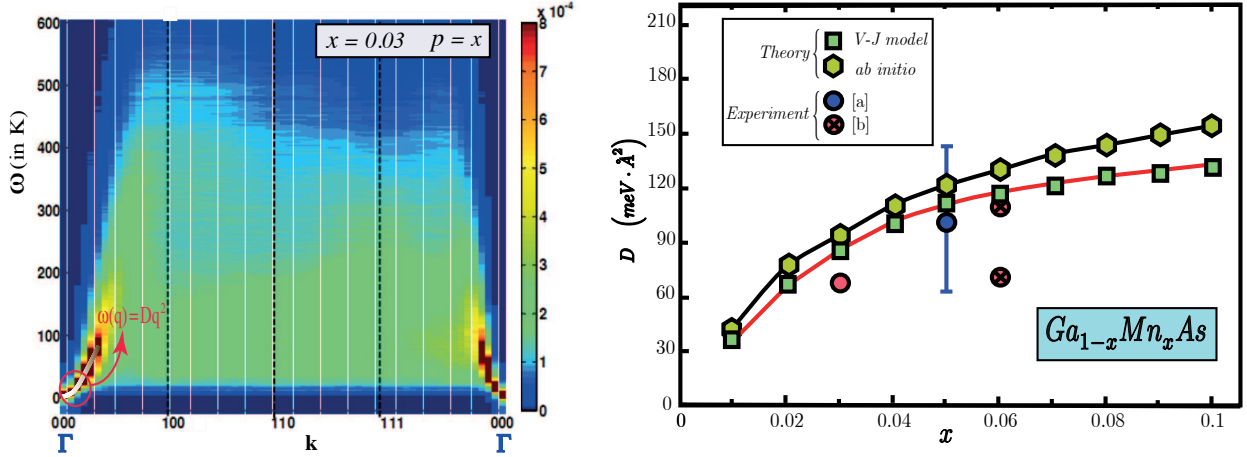}}
\caption{
(Color online) 
(Color online)((a) Left) Intensity plot of the average dynamical spectral function $\bar{A}({\bf q},\omega)$ (see text) in the ($\bf k$,$\omega$) plane for a 3$\%$  Mn doped GaAs compound and a hole density $p=x$ (see Ref. \onlinecite{georges2007}). The couplings used have been obtained from LSDA calculations. ((b) Right) Spin stiffness D (meV \AA$^2$) in Ga$_{1-x}$Mn$_{x}$As as a function of $x$ (see Ref. \onlinecite{akash2011}) . The hole concentration is set to $p=x$ (well annealed compounds). Squares correspond to the V-J model calculations \cite{akash2011}), hexagons to ab initio values  \cite{georges2007}, (a) and (b) circles to experimental measurements extracted from Ref. \onlinecite{goennenwein2003} and \onlinecite{sperl2008}. The circles with a cross correspond to annealed samples.
}
\label{fig6}
\end{figure}

%\begin{figure}[htbp]
%\centering
%\subfigure{
%\includegraphics[width=3.1in,angle=0]{Figv6-left.eps}
%\label{fig6a}
%}
%\hspace{-0.70cm}
%\subfigure{
%\includegraphics[width=3.2in,angle=0]{Figv6-right.eps}
%\label{fig6b}
%}
%\label{fig6}
%\caption{
%}
%\end{figure} 

We now discuss the low energy spin excitation spectrum in III-V doped systems. We will focus on the case of Mn doped GaAs since, to our knowledge, no experimental data are available for the other compounds. For this purpose we have calculated the dynamical spectral function that provides deeper insight 
into the underlying spin dynamics. This physical quantity can be directly and accurately probed by inelastic neutron scattering (INS) experiments. 
The averaged dynamical spectral function is defined as follows $\bar{A}({\bf q},\omega) \equiv \langle A({\bf q},\omega)\rangle_c$ where $\langle ... \rangle_c$ means average over disorder. For a given configuration the spectral function is given by,
\begin{eqnarray}
A({\bf q},\omega) =  \sum_\alpha A_{\alpha}({\bf q})\delta(\omega-\omega_{\alpha})
\label{Aqw}
\end{eqnarray}
where,
\begin{eqnarray}
 A_{\alpha}({\bf q})= \frac{1}{N_{imp}} \sum_{ij} \lambda_j \langle i|\Psi_\alpha^{R}\rangle \langle\Psi_\alpha^{L}|j\rangle e^{i\bf q(r_i-r_j)}
\label{Acq}
\end{eqnarray}
In Fig.~\ref{fig6}-left we have plotted $\bar{A}({\bf q},\omega)$ in the  ($\bf k$,$\omega$) plane. The calculations are performed for a Mn concentration of 3$\%$ in GaAs. The Mn-Mn couplings are those obtained from local spin density approximation calculations (see Ref. \onlinecite{georges2007}). In contrast to what one usually observes in weakly disordered magnetic
systems, in the dilute Mn doped GaAs, well-defined excitations exist only in a restricted region of the
Brillouin zone centered around the $\Gamma$-point (q =(0, 0, 0)). As we move away from  the center of the Brillouin zone, the width of the excitation increases rapidly. Beyond a momentum cut-off q$_c$
no well defined excitations exists anymore. 
This  is a consequence of the short range  nature of the exchange integrals.
A recent study \cite{zrmno2-epl} of the magnetic properties of Mn doped ZrO$_2$ has revealed that $q_c =A(x-x_c)^{1/3}$, where $x_c$ is the percolation threshold. That should also hold in the present case.
Note that $x_c$ is about 0.0075 in Mn doped GaAs (see Fig.~\ref{fig4}). In the region of well defined excitations we find the expected quadratic magnon dispersion $\omega$(q)=D$(x)$q$^2$ , where D is the so called spin stiffness.
In Fig.~\ref{fig6}-right we have plotted D as a function of $x$. The theoretical values, obtained both via V-J model and ab initio, are shown together with available experimental data. Details concerning the calculations are given in Refs. \onlinecite{georges2007} and \onlinecite{akash2011}. First, we observe, for the whole concentration range a very good agreement between the V-J and ab initio based calculations. We have also found a good quantitative agreement with the experimental data for both 3$\%$ and 5$\%$ doped samples. Regarding the 6$\%$  doped case, one of the annealed sample agree very well whilst the other has a lower value.  In the latter case the average experimental spin stiffness is 90 $\pm$ 20 meV $\AA^2$ . On the other hand the V-J model  and ab initio calculations give respectively 120 $\pm$ 30 meV$\AA^2$ and 130$\pm$ 30 meV$\AA^2$. Note that, the uncertainty in the theoretical values  result from the sensitivity to the magnetic couplings at relatively high distances between localized spins. Thus the agreement is still reasonably good for this concentration as well.

\section{Conclusions}

We have demonstrated that we can describe quantitatively the physics of DMS, both transport and magnetism, within a coherent picture. The minimal V-J Hamiltonian is the missing link that bridges the gap between complex and material specific first principle studies and model approaches. It has been shown that the physics is essentially controlled by the position of the p-d acceptor level with respect to the top of the valence band. This model approach continuously explain the change in the nature of the couplings. The two extreme regimes, RKKY in Mn doped II-VI such as (Zn,Mn)Te and double exchange like in (Ga,Mn)N, are described within the same picture. The agreement between ab initio, V-J model and experimental data are impressive (Curie temperatures, low energy magnetic excitation spectrum). The V-J model clearly explains the reason why Mn doped GaAs exibits the highest critical temperature among both II-VI and III-V compounds and its proximity to metal-insulator transition. Hence, this model provides an efficient tool to find other pathways towards room temperature ferromagnetism, such as the influence of correlated disorder and nanostructuration of the materials for instance.


\begin{thebibliography}{99}
%\bibliography{apssamp}

\bibitem{junqwirth} T. Jungwirth et al., Rev. Mod. Phys. \textbf{78,} 809 (2006).
\bibitem{satormp} For a review see K. Sato et al. Rev. Mod. Phys. \textbf{82,} 1633 (2010).
\bibitem{timm} C. Timm, J. Phys. Condens. Matter \textbf{15,} R1865 (2003).
\bibitem{dietl} T. Dietl et al. Science. \textbf{287,} 1019 (2000).
\bibitem{kl} J. M. Luttinger and W. Kohn, Phys. Rev. \textbf{97,} 869 (1955). 
\bibitem{kane} E. O.  Kane, J. Phys. Chem. Solids \textbf{1,} 249 (1957).

\bibitem{wierzbo} M. Wierzbowska et al., Phys. Rev. B \textbf{70,} 235209 (2004)

\bibitem{burch} K.S. Burch et al. ,Phys. Rev. Lett. \textbf{97,} 087208 (2006). 
\bibitem{singley1} E.J. Singley et al., Phys. Rev. Lett. \textbf{89,} 097203 (2002). 
\bibitem{singley2} E.J. Singley et al., Phys. Rev. B \textbf{68,} 165204 (2003). 

\bibitem{hayashi} T. Hayashi et al., Appl. Phys. Lett. \textbf{78,} 1691 (2001)
\bibitem{matsukura} F. Matsukura et al., Phys. Rev. B \textbf{57,} 2037 (1998)
\bibitem{yazdani} A. Richardella et al.  Science \textbf{327,} 665 (2010).




\bibitem{dieudonne1953} More details on bi-orthogonality can be found in J. Dieudonn\'e, Mich. Math. J. \textbf{2,} 7 (1953).
\bibitem{georges2005epl} G. Bouzerar et al. Appl. Phys. Lett. \textbf{85,} 4941 (2004); G. Bouzerar et al. Europhys. Lett. \textbf{69,} 812 (2005).
\bibitem{georges2007} G. Bouzerar, Eur. Phys. Lett. \textbf{79,} 57007 (2007).
\bibitem{akash2010} A. Chakraborty and G. Bouzerar, Phys. Rev. B \textbf{81,} 172406 (2010). 
\bibitem{gb-prb-compens} G. Bouzerar et al. ,Phys. Rev. B \textbf{72,} 125207 (2005). 
\bibitem{rb-prb-compens} R. Bouzerar et al. , Phys. Rev. B \textbf{82,} 035207 (2010). 
\bibitem{gb-de-2007} G. Bouzerar and O. C\'epas  Phys. Rev. B \textbf{76,} 020401 (2007).
\bibitem{gb-prb-canting} G. Bouzerar, R. Bouzerar and O. C\'epas  Phys. Rev. B \textbf{76,} 144419 (2007).
\bibitem{MC1} L. Bergqvist et al., Phys. Rev. Lett. \textbf{93,} 137202 (2004).
\bibitem{MC2} K. Sato et al. , Phys. Rev. B \textbf{70,} 201202 (2004). 
\bibitem{richard-epl2007} R. Bouzerar et al. , Eur. Phys. Lett. \textbf{78,} 67003 (2007). 
\bibitem{richard1} R. Bouzerar and G. Bouzerar, Europhys. Lett. \textbf{92,} 47006 (2010). 
\bibitem{barthel} S. Barthel et al., Eur. Phys. J. B  \textbf{86,} 11 (2013).
\bibitem{transport} R. Bouzerar and G. Bouzerar, New Journal of Physics \textbf{13,} 023002 (2011).


\bibitem{edmonds1} K.W. Edmonds et al. Appl. Phys. Lett \textbf{81,} 4991 (2002).
\bibitem{edmonds2} K. W. Edmonds et al. Phys. Rev. Lett. \textbf{92,} 03201 (2004).
\bibitem{chiba} D.  Chiba et al. Applied Physics Letters \textbf{82,}, 3020 (2003).
\bibitem{jungwirth} T. Jungwirth et al., Phys. Rev. B \textbf{72,} 165204 (2005).
\bibitem{stone} P. R. Stone et al. Phys Rev. Lett. \textbf{101,} 087203 (2008).

\bibitem{akash2011} A. Chakraborty, R. Bouzerar, and G. Bouzerar, Eur. Phys. J. B \textbf{81,} 405 (2011). 
\bibitem{goennenwein2003} S.T.B. Goennenwein et al., Appl. Phys. Lett. \textbf{82,} 730 (2003).
\bibitem{sperl2008} M. Sperl et al., Phys. Rev. B \textbf{77}, 125212 (2008).

\bibitem{rb-prb-compens} R. Bouzerar et al. , Phys. Rev. B \textbf{82,} 035207 (2010). 
\bibitem{zrmno2-epl} A Chakraborty and G. Bouzerar, Eur. Phys. Lett. \textbf{104,} 57010 (2013).








 \end{thebibliography}
\end{document}